%% file: main.tex
\RequirePackage[hyphens]{url}

\documentclass[sigconf]{acmart}

\usepackage{xcolor}
\usepackage{booktabs}
\usepackage{graphicx}
\usepackage{subfiles}
\usepackage{balance}
\usepackage{todonotes}

\renewcommand\footnotetextcopyrightpermission[1]{} 

\graphicspath{{images/}}

\setcopyright{none}

\acmDOI{}

\acmISBN{}

\acmConference[JCDL 2019]{ACM/IEEE Joint Conference on Digital Libraries}{June 2019}{Urbana-Champaign, IL, USA}
\acmYear{}
\copyrightyear{}

\begin{document}
\begin{NoHyper}

\title[Do Authors Deposit on Time? Tracking Open Access Policy Compliance]{Do Authors Deposit on Time?\\ Tracking Open Access Policy Compliance}

\author{Drahomira Herrmannova}
\orcid{0000-0002-2730-1546}
\affiliation{%
  \institution{Knowledge Media Institute\\ The Open University}
  \city{Milton Keynes}
  \country{United Kingdom}
}
\email{orcid.org/0000-0002-2730-1546}
\email{dasha.herrmannova@open.ac.uk}

\author{Nancy Pontika}
\orcid{0000-0002-2091-0402}
\affiliation{%
  \institution{Knowledge Media Institute\\ The Open University}
  \city{Milton Keynes}
  \country{United Kingdom}
}
\email{orcid.org/0000-0002-2091-0402}
\email{nancy.pontika@open.ac.uk}

\author{Petr Knoth}
\orcid{0000-0003-1161-7359}
\affiliation{%
  \institution{Knowledge Media Institute\\ The Open University}
  \city{Milton Keynes}
  \country{United Kingdom}
}
\email{orcid.org/0000-0003-1161-7359}
\email{petr.knoth@open.ac.uk}

\renewcommand{\shorttitle}{Do Authors Deposit on Time? Tracking Open Access Policy Compliance}

\begin{abstract}
Recent years have seen fast growth in the number of policies mandating Open Access (OA) to research outputs. We conduct a large-scale analysis of over 800 thousand papers from repositories around the world published over a period of 5 years to investigate: a) if the time lag between the date of publication and date of deposit in a repository can be effectively tracked across thousands of repositories globally, and b) if introducing deposit deadlines is associated with a reduction of time from acceptance to public availability of research outputs. We show that after the introduction of the UK REF 2021 OA policy, this time lag has decreased significantly in the UK and that the policy introduction might have accelerated the UK's move towards immediate OA\footnote{The term ``immediate OA'' is usually used to refer to outputs that are available immediately upon publication without any embargo periods. In this context, we do not consider embargoes and use it simply to mean availability upon publication.
} compared to other countries. This supports the argument for the inclusion of a time-limited deposit requirement in OA policies. 
\end{abstract}

%
%



\maketitle

\section{Introduction}
\label{sec:introduction}
\subfile{sections/01_introduction}

\section{Related Work}
\label{sec:related_work}
\subfile{sections/02_related_work}

\section{Methodology}
\label{sec:background}
\subfile{sections/03_methodology}

\section{Data Preparation}
\label{sec:data}
\subfile{sections/04_data}

\section{Results}
\label{sec:experiments}
\subfile{sections/05_experiments}

\section{Discussion}
\label{sec:discussion}
\subfile{sections/06_discussion}

\section{Conclusion}
\label{sec:conclusion}
\subfile{sections/07_conclusion}

\bibliographystyle{acmbibformat}
\bibliography{bibliography}

\clearpage

\appendix

\subfile{sections/99_appendix}

\end{NoHyper}
\end{document}

%% file: sections/01_introduction.tex
More than seventeen years have passed since the definition of Open Access (OA) has been agreed \cite{boai}. OA, which refers to scientific literature that is online and available free of cost to the end user, questions the traditional publishing business model relying on paywalls and advocates for a shift towards alternative, more cost-effective publishing models delivering free access to research outputs for all \cite{sample_2012_harvard,shieber_2013_open,library_budget,suber_2003_taxpayer}. These arguments have been gradually influencing researchers, research organisations, and funders, resulting in the creation of new OA policies. As of January 2019, according to the Registry of Open Access Repository Mandates and Policies\footnote{\url{https://roarmap.eprints.org/}}, there are 732 institutional and 85 funder OA policies globally.

OA policies provide authors with criteria 
for making their research outputs available as OA \cite{picarra_2015_monitoring}. These criteria typically include when and where should the research outputs be deposited or published and what version of the manuscript (e.g. pre-print vs. post-print) should be made openly accessible \cite{picarra_2015_monitoring}. Arguably one of the most significant OA policies, the UK Research Excellence Framework (REF) 2021 Open Access Policy\footnote{\url{https://www.ref.ac.uk/about/what-is-the-ref/}}, was introduced in the UK in March 2014 \cite{hefce_ref2021_oa}. The significance of this policy lies in two aspects: 1) the requirement to make research outputs OA is linked to performance review, creating a strong incentive for compliance \cite{xia_2012_review,swan_2015_working,vincentlamarre_2016_estimating}, and 2) it affects over 5\% of global research outputs\footnote{According to Scimago Journal \& Country Rank (\url{https://www.scimagojr.com/countryrank.php?year=2017}), in 2017 the UK was the third largest producer of research outputs, representing 5.42\% of global research outputs. It ranked third after the US with 17.71\% and China with 14.38\% percent share.}. Under this policy, only compliant research outputs will be evaluated in the national Research Excellence Framework. Over 52 thousand academic staff from 154 UK universities submitted over 190 thousand research outputs in the most recent REF (2014) \cite{ref2014_key_facts}. The UK REF 2021 OA policy is not the only major nation-wide development -- the U.S. Public Access Plan \cite{public_access_plan} introduced in 2013 and the European Commission supported ``Plan S'' \cite{plan_s}, are just two more examples of a global shift towards Open Access.

\textbf{The problem.} The growth of OA and the introduction of new policies, such as the REF 2021 Open Access Policy, 
has brought forth important questions and implications, some universal and some policy-specific. Even when authors deposit their work in OA repositories, does this happen immediately, or is the deposit delayed? What effect does the introduction of policies have on the practice of publishing OA? Is there evidence to support that introducing OA policies reduces the time from acceptance to the open availability of research outputs? More importantly, how can compliance with OA policies be tracked, particularly when specific time-frames for making research outputs OA are in place? While recent studies analysing compliance with OA policies \cite{lariviere_2018_authors,khoo_2018_embargo} and the prevalence of OA \cite{piwowar_2018_state} have focused on whether articles are eventually made openly available, they have not taken into consideration the time lag between the acceptance/publication of an article and its online availability (deposit into an OA repository). Two existing studies which have taken deposit dates into consideration \cite{swan_2015_working,vincentlamarre_2016_estimating} are now outdated, are not easily reproducible, and have not used these dates to assess compliance (i.e. to understand whether authors deposit on time in accordance with existing policies) but instead used these dates to study policy effectiveness (i.e. to understand whether certain types of policies shorten the time between publication and deposit). If we can measure the time lag between publication and deposit, can we assist authors and institutions in improving their compliance with OA policies?

\textbf{Research questions.} In this paper we analyse the time lag between article publication dates and dates of their deposit into OA repositories. We will further refer to the time lag between these dates simply as \textit{deposit time lag}\footnote{Existing studies sometimes refer to the difference between the publication and deposit dates as ``deposit latency'' \cite{swan_2015_working,vincentlamarre_2016_estimating}. However, because the term ``latency'' is in computer science typically associated with a different meaning, we chose to use the term ``time lag'' instead.}. We analyse deposit time lag across country, time, repository, and  discipline. Furthermore, we investigate whether introducing a mandatory policy in the UK -- the REF 2021 Open Access Policy, which requires depositing research outputs within a specific period -- affected this time lag. 

To study deposit time lag and compliance with the policy, we use data from Crossref\footnote{\url{https://www.crossref.org/}}, the largest Digital Object Identifier (DOI) registration agency, and from CORE\footnote{\url{https://core.ac.uk/}}, the largest full text aggregation service collecting OA research outputs from institutional and subject repositories and from journals around the world \cite{Knoth_2012}. 
After matching article metadata from Crossref and from CORE 
we analyse the time lag between publication dates we receive from Crossref and deposit dates we receive from CORE. 
Using this data, we answer the following research questions:

\begin{enumerate}
    \item How does deposit time lag vary across time, country, institution, and discipline?
    \item What proportion of UK research outputs was not deposited on time to comply with the REF 2021 OA Policy?
    \item Is the REF 2021 OA policy affecting how soon are publications made OA?
    \item How does the change in the deposit time lag in the UK over the past several years compare to other countries?
\end{enumerate}

\textbf{Findings.} We show that the time between publication and deposit has globally significantly decreased. We also show that while there are notable differences in deposit time lag of different subjects, there are even larger differences between different institutions, even when considering only publications from the same discipline. This suggests institutions may be stronger drivers of OA than discipline culture. Furthermore, we show the introduction of the UK REF OA Policy might have accelerated the UK's move towards immediate OA compared to other countries.

\textbf{Contributions.} We present a method for automated tracking of deposit time lag which can be applied to research outputs world-wide. Using this method, we provide the first large scale analysis of deposit time lag. Ours is also the first study to quantitatively analyse deposit time lag in relation to the REF 2021 OA Policy. Our results support the argument for the inclusion of a time-limited deposit requirement in OA policies. Finally, to support further studies on the deposit of research outputs into OA repositories, we release our dataset of 800 thousand publications and the source codes of our analysis\footnote{\url{https://github.com/oacore/jcdl_2019}}.

\textbf{Outline.} This paper is organised as follows. First, in Section \ref{sec:related_work} we review previous work related to our study. Next, in Section \ref{sec:background} we describe our data collection process and the methodology used in our analysis. In Section \ref{sec:data} we explain how we prepare our dataset, and in Section \ref{sec:experiments} we present the results of our analysis. Finally, Section \ref{sec:discussion} discusses limitation of the present work and future goals.

%% file: sections/02_related_work.tex
In this section we discuss work related to our research. In particular, we focus on two topics: 1) studies that try to estimate the proportion of all research publications that are openly accessible and 2) studies that analyse compliance with specific OA policies. We close this section by discussing the differences between our study and previous work. 

Particularly in recent years many studies have been conducted that have tried to estimate the proportion of existing research that is available as OA \cite{bjork_2010_open,gargouri_2012_green,archambault_2013_proportion,archambault_2014_proportion,khabsa_2014_number,piwowar_2018_state,lariviere_2018_authors}. While an earlier study identified OA articles using manual Google search \cite{bjork_2010_open}, the later studies use automated methods based on web crawling \cite{gargouri_2012_green,khabsa_2014_number}, database searching \cite{piwowar_2018_state,lariviere_2018_authors}, or a combination of both \cite{archambault_2013_proportion,archambault_2014_proportion}. One of the two most recent studies has estimated the proportion of OA articles to be at least 28\% overall (a finding similar to \cite{gargouri_2012_green,khabsa_2014_number}), with 45\% of articles published in 2015 
being OA \cite{piwowar_2018_state}. The most recent study we know of \cite{lariviere_2018_authors} has utilised the same method as \cite{piwowar_2018_state}, but focused on publications subject to OA policies of selected funders, revealing that two thirds of these publications were available as OA.

Two of the studies \cite{gargouri_2012_green,lariviere_2018_authors} are of particular interest because they investigated the proportion of OA articles in relation to specific policies. \citeauthor{gargouri_2012_green} \cite{gargouri_2012_green} have demonstrated that the proportion of OA articles at institutions with OA policies was three times as high as at institutions without them. Interestingly, the study has also shown not all articles were made available online upon publication but were instead deposited retrospectively. \citeauthor{lariviere_2018_authors} \cite{lariviere_2018_authors} investigated twelve funders (the European Research Council and eleven funders from the UK, US and Canada) which implemented OA policies. The study has revealed significant differences in the proportion of OA publications between different funders, even when considering funders from the same discipline. In particular, funders which required depositing into a repository upon publication had significantly higher proportion of OA articles than funders which allowed deposit after publication. While the authors have observed differences between disciplines, finding significant variations between funders within the same discipline has led the authors to conclude the funding agency may be a stronger driver of OA publishing than the culture within a discipline.


The above mentioned studies look at how many publications are available as OA compared to how many publications appear behind paywalls. However, as \citeauthor{gargouri_2012_green} \cite{gargouri_2012_green} have indirectly shown, the open online availability of a publication does not necessarily ensure compliance with a given policy. A number of policies, including the UK REF 2021 OA Policy and the US National Institutes of Health (NIH) Public Access Policy, require deposit by a certain date -- three months after acceptance in the case of the REF 2021 OA Policy and upon publication in the case of the NIH Public Access Policy. The approach utilised by the above mentioned works would typically mean even publications which were deposited retrospectively could be considered compliant with these two policies.

Only a handful of studies have investigated specific details of existing policies \cite{vincentlamarre_2016_estimating,swan_2015_working,khoo_2018_embargo}. \citeauthor{vincentlamarre_2016_estimating} \cite{vincentlamarre_2016_estimating} analysed research articles published 
by 67 institutions with an OA mandate, i.e. an OA policy which was mandatory rather than recommended. The studied mandates were broken down into eight specific conditions such as deposit timing and embargo length, and the study investigated how these conditions relate to mandate compliance. They found that one value for three of the eight conditions (immediate deposit required, deposit required for performance evaluation, unconditional opt-out allowed for the OA requirement but no opt-out for deposit requirement) was strongly associated with higher deposit rates as well as with lower deposit time lag. \citeauthor{swan_2015_working} \cite{swan_2015_working} have conducted a similar study and compared specific policy conditions with deposit rates and time lag for 122 institutions with mandatory OA policies. Similarly as in the case of \cite{vincentlamarre_2016_estimating}, the authors have identified three criteria which were associated with improved deposit rates (deposit mandatory, deposit cannot be waived, deposit should be linked with research evaluation). \citeauthor{khoo_2018_embargo} \cite{khoo_2018_embargo} have focused on embargo periods and studied the rate at which neuroscientists in Australia and Canada publish in journals with embargo periods that are not compliant with funder policies, i.e. are longer than 12 months. Interestingly, they observed no reduction in the number of articles published in journals with non-compliant embargo periods after new funder policies were introduced in Australia and Canada, despite these policies being mandatory.

In the present work we investigate how much time does it take for authors to deposit their articles in OA repositories in relation to when these articles get published. 
Our work differs from the aforementioned studies in a number of ways. In contrast to \cite{vincentlamarre_2016_estimating} and \cite{swan_2015_working} who correlated deposit time lag with specific policy conditions, we instead analyse how deposit time lag differs across a number of dimensions such as country and discipline. We also address what we envision as a future step in assisting the OA movement -- automated and reproducible tracking of policy compliance. By utilising the CORE aggregator which harvests content from thousands of repositories globally, we are able to study how many publications get deposited in multiple places and whether recognising these multiple copies can enable faster access to research. Ours is also the first study to quantitatively analyse the UK REF 2021 OA Policy.

%% file: sections/03_methodology.tex
In this section, we describe the datasets and the methodology used to answer our research questions. As one of the aims of this work is to study compliance with the UK REF 2021 OA Policy, we start by introducing the policy.

Compliance with the REF 2021 OA Policy is met when authors deposit (self-archive) the post-print (also called the ``author accepted manuscript,'' i.e. author's final version of the manuscript where all the peer review suggestions have been addressed but without the publisher's typesetting) into an institutional or a subject repository within three months from the acceptance of the publication \cite{hefce_ref2021_oa,swan_2014_hefce}. 
The policy affects journal articles and conference proceedings with an International Standard Serial Number (ISSN), which constitute the majority (77\%) of outputs submitted to the latest REF \cite{kerridge_2014_open}. Although the policy was introduced in 2014, the implementation period started in April 2016 to allow universities to create the necessary infrastructure for tracking compliance.

To collect the data needed for the analysis of deposit time lag world-wide, we use the following data sources:

\begin{itemize}
    \item \textbf{Crossref}\footnote{\url{https://www.crossref.org/}} is the largest DOI registration agency. Crossref stores publication metadata associated with each DOI that is registered with the service. At the time of writing, Crossref contained 103 million records\footnote{\url{https://www.crossref.org/dashboard/}}.
    \item \textbf{CORE}\footnote{\url{https://core.ac.uk/}} is the world's largest OA aggregation service \cite{core_largest}, collecting OA research outputs from institutional and subject repositories\footnote{Subject repositories aggregated by CORE include e-print repositories such as ArXiv which is often used to deposit pre-prints 
    as well as post-prints. 
    The latest REF 2021 submission guidelines state e-print repositories will be considered acceptable for compliance purposes \cite{re_ref2021}. We have therefore included these repositories in our analysis.} and from journals worldwide \cite{Knoth_2012}. As such, CORE provides a single interface for accessing data from repositories around the world. At the time of writing, CORE aggregated content from over 3,700 repositories and contained 135 million article records. While there are other services such as OpenAIRE 
    and BASE
    , which aggregate data from repositories; OpenAIRE has an order of magnitude smaller dataset (25 million records) and neither BASE nor OpenAIRE make the datasets publicly available for download and analysis. Furthermore, judging from the user interfaces of both, deposit dates do not appear to be available.
\end{itemize}

\begin{figure}
  \includegraphics[width=0.9\linewidth]{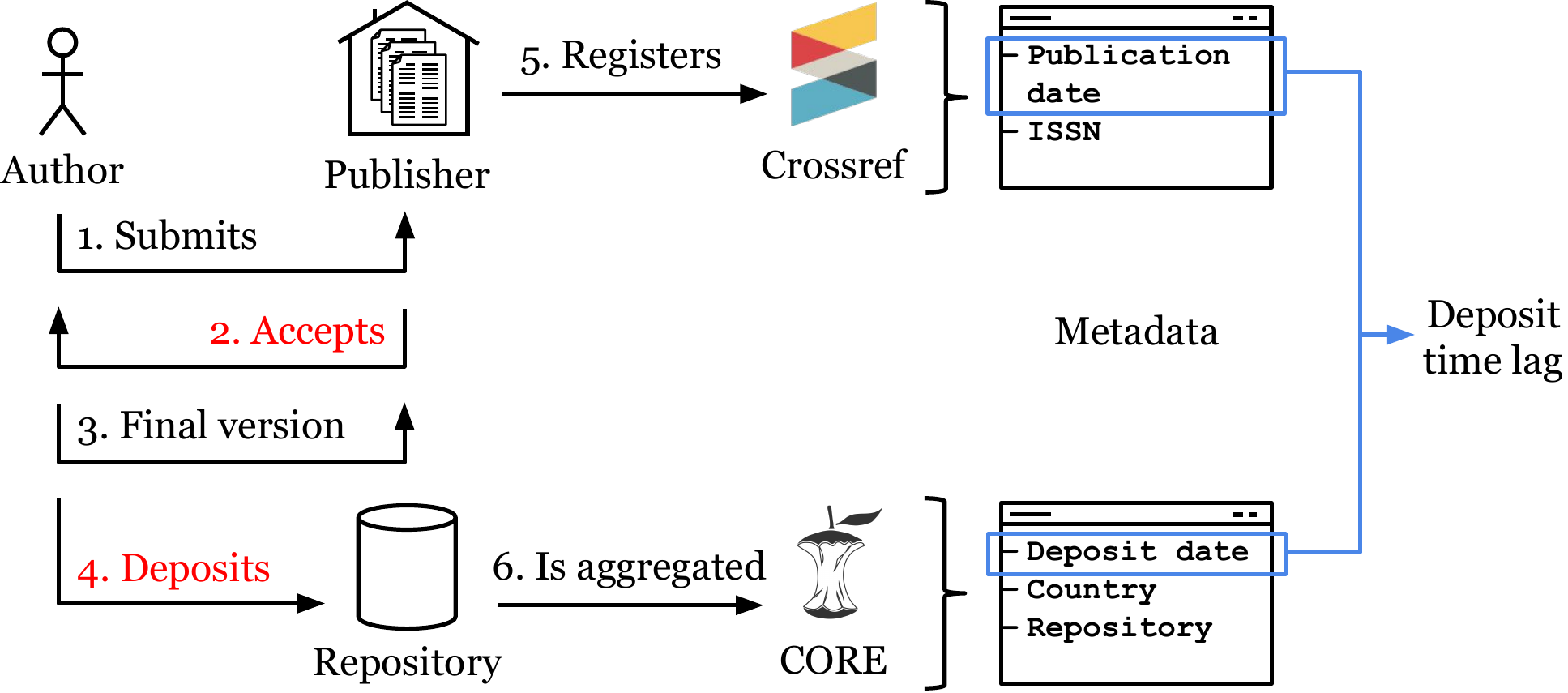}
  \caption{A visual depiction of the publishing and data collection process, which is started by a submission and an acceptance of a publication. Steps mentioned in the REF 2021 OA Policy are shown in red. The dates we acquire from the two databases and use to calculate the deposit time lag are highlighted with a blue frame.}
  \label{fig:data}
\end{figure}

Figure \ref{fig:data} shows Crossref and CORE along with the data they collect and depicts the process of how published articles get entered into these systems. The process is started when an author submits and a publisher accepts a manuscript. The REF 2021 Open Access Policy stipulates that the author's final version of the manuscript (i.e. the post-print) must be deposited into a repository within three months of acceptance. The acceptance and deposit steps, which are mentioned in the policy, are shown in red in the figure.

Upon receiving the author's final version of the manuscript, the publisher registers this manuscript with Crossref. Crossref then stores metadata associated with the publication, including the date of publication. Furthermore, once the author's final version of the manuscript is deposited in a repository, the metadata of the publication including the date it was deposited into the repository is propagated into CORE through its aggregation service.

The REF 2021 OA Policy requires papers to be deposited into a repository within a certain time frame relative to the date of acceptance. However, when the policy was introduced, the date of acceptance was not tracked by Crossref or by most repositories and other databases. Although Crossref metadata now contain an \textit{accepted} field, this field is only populated for a small fraction of publications (this will be further discussed in Section \ref{sec:data:acceptance_date}). Furthermore, while repositories have since the introduction of the policy created infrastructure for recording the acceptance date, the date is unlikely to be available for publications published prior to the policy taking effect and for non-UK publications. Consequently, the acceptance date does not allow us to study compliance with the policy over time or compare the UK to other countries. Therefore, to measure deposit time lag and non-compliance with the policy, we use dates of publication instead of acceptance dates. 

\subsection{Data}
\label{sec:methodology:data}

As mentioned above, we use Crossref and CORE to collect data for our analysis. More specifically, we use Crossref to obtain publication dates and ISSN numbers, and CORE to obtain deposit dates, repository names, and for institutional repositories also locations (specifically the country of the repository). 

Additionally, to ensure correct deposit dates for older documents, we have applied the following procedure. CORE harvests documents from repositories using the Open Archives Initiative Protocol for Metadata Harvesting\footnote{https://www.openarchives.org/pmh/} (OAI-PMH). The OAI-PMH metadata do not contain a deposit date field, but only a last update field. Thus, the last update field will contain a deposit date of an article up until the article's metadata is updated in the repository. The metadata does not distinguish which version of the article is presented. In September 2018, CORE created infrastructure which allows it to store the first date it receives as the deposit date and any subsequent dates as dates of updates. To ensure correct deposit dates for documents deposited prior to September 2018, we have created web scrapers for the following repositories: repositories using DSpace, EPrints, or Invenio software, and additional individual scrapers for ArXiv and Zenodo. The choice of repositories we created scrapers for was made based on a) availability of deposit dates on the website and b) whether we were able to match a repository page URL to a specific OAI-PMH metadata record.

Furthermore, we used Mendeley\footnote{https://www.mendeley.com/} to obtain information about publications' subjects using the profiles of those who read the publications. 
Mendeley is a reference manager that can be used to manage a research library and provides an API that can be queried to obtain information about how many people have added a certain publication in their libraries. When users create Mendeley accounts, they are asked about their fields of study. We have used the information about how many users from each field of study have bookmarked a certain publication to categorise publications into subject categories. The details of how we did this are described in Section \ref{sec:data:subject_distrib}.

\subsection{Compliance categories}
\label{sec:methodology:compliance_categories}

Based on the available data, for the analysis of the REF 2021 OA Policy we can assign each publication to one of the following \emph{compliance categories}:

\subsubsection{Definitely non-compliant:} a publication has been deposited into a repository and its first date of deposit is later than three months after its original date of publication. 
This category may not include all non-compliant publications as some may fall into the ``likely compliant'' category below, depending on their actual date of acceptance. However, using this classification, we can be certain that all publications within the non-compliant category are indeed non-compliant, i.e. this category will have 100\% precision but not 100\% recall.


\subsubsection{Likely compliant:} a publication has been deposited into a repository and its deposit date is within a three months period of its original publication date or earlier. 
This category may include some non-compliant publications, depending on the actual date of acceptance. However, given the way it's defined, we can be certain that all truly compliant publications will fall into this category, i.e. this category will have 100\% recall but not 100\% precision.

%% file: sections/04_data.tex

We started by obtaining a complete data dump from Crossref and CORE. 
Our Crossref data dump was obtained in May 2018 and our CORE dump in March 2019 (the reason why our CORE dump was obtained later was to allow enough time for publications to be deposited and aggregated by CORE). We then filtered out all documents with a missing title, year of publication, or author names. Additionally, we filtered out any Crossref documents where the metadata contained only the year of publication but not the month of publication. If a day of publication, but not the year or month, was missing, we used the first day of the month as the day when the paper was published, e.g. if we knew a paper was published in 2017-09, we replaced the date with 2017-09-01. Finally, we removed all documents from both datasets which were published prior to 2013. 
After this filtering we were left with 18,753,649 CORE articles and 15,832,311 Crossref articles.

Title, year of publication, and the last name of the first author were then used to merge the two datasets. As not all documents in CORE contain a DOI, we were unable to use DOIs to match documents between Crossref and CORE. On the other hand, title, author, and year information are available for most documents. Matching documents by title, year, and first author name is a strict approach which results in lower recall, because authors may not be listed in the correct order, different spelling or hyphenation of the titles and author names may be used, etc. However, this approach produces cleaner and more reliable data (the accuracy of this matching method is 95.27\%, a more detailed analysis of the accuracy is provided in Section \ref{sec:matching}) and for the purposes of the analysis this was our aim.

Titles and author names were cleaned by removing all characters other than alphanumeric characters and underscores, and by converting all characters to lowercase. Additionally, we have normalised the text by replacing accented characters and special characters appearing in non-English alphabets with their non-accented/English versions (e.g. by converting ``Fran\c{c}ois'' to ``Francois''). The data was then merged using exact match on the title, year of publication, and last name of the first author. Because one article can be deposited in multiple repositories (for example if the authors of the article are affiliated with different institutions and all deposit the article in their respective repositories), we have additionally grouped all CORE articles that were matched to the same Crossref article into one record using Crossref DOI. This grouping reduces the size of the dataset by about half a million records and the merged and grouped dataset contains 1,589,469 rows. Finally, we have used our repository scrapers (Section \ref{sec:methodology:data}) to obtain correct deposit dates. We were able to obtain deposit dates for 808,984 documents in our dataset. Table \ref{tab:dataset_size} shows the final dataset size.



\bgroup
\begin{table}
\caption{Dataset size.}
\label{tab:dataset_size}
\begin{tabular}{|l|r|}
\hline 
Unique CORE articles           & 948,044 \\
Unique Crossref articles       & 808,984 \\
Links between Crossref \& CORE & 985,175 \\
\hline 
\hline
Final dataset size (after grouping) & 808,984 \\
\hline 
\end{tabular}
\end{table}
\egroup

\subsection{Analysis of our matching method}
\label{sec:matching}

As the results of our analysis are impacted by the above mentioned matching method, we need to be confident that the accuracy of the matching is high. To measure this accuracy, we compare DOIs between all pairs of matched documents. There are 985,175 document pairs in total (Table \ref{tab:dataset_size}) out of which 354,897 don't have a DOI in CORE (36.02\%). Of the remaining 630,278 that have a DOI both in Crossref and in CORE, 595,202 have exactly matching DOIs (94.43\% of the 630 thousand pairs) and 35,076 have DOIs that do not match (5.57\%). 

We have investigated the non-matches and observed that it is often because of minor differences which seem like errors introduced during the deposit in the repository. More specifically, DOIs obtained from CORE often have additional text appended at the end (Table \ref{tab:doi_diffs}, Example 1) while clearly referring to the same document. This is not the case for the opposite scenario, 
as CORE DOIs with missing characters can often match multiple Crossref DOIs (Table \ref{tab:doi_diffs}, Example 2). There are 5,264 DOI pairs (15.01\% of the non-matching DOI pairs) where Crossref DOI is substring of the CORE DOI, i.e. CORE DOI contains additional characters. If we consider these as correct matches, the accuracy of the matching method is 95.27\%.

\bgroup
\begin{table}
\caption{Examples of differences between DOIs obtained form Crossref and from CORE.}
\label{tab:doi_diffs}
\begin{tabular}{|l|l|}
\hline 
\multicolumn{2}{|l|}{Example 1} \\
\hline 
CORE DOI & 10.1002/2016jd026252/abstract \\
Crossref DOI & 10.1002/2016jd026252 \\
\hline 
\hline 
\multicolumn{2}{|l|}{Example 2} \\
\hline 
CORE DOI & 10.1088/0031-8949 \\
Crossref DOI 1 & 10.1088/0031-8949/2013/t156/014026 \\
Crossref DOI 2 & 10.1088/0031-8949/90/9/095101 \\
\hline
\end{tabular}
\end{table}
\egroup

Given the 95.27\% matching accuracy, we estimate that 338,110 document pairs, which do not have a DOI in CORE, were matched correctly. If we were to match documents by DOIs instead, we would have missed these. Furthermore, evaluating the accuracy of the method would have been more time consuming (it would require a manual check) and would likely be less precise.

\subsection{Repository distribution}

We are interested in studying the differences in deposit time lag at different institutions. However, Crossref only contains affiliation information for a small subset of the publications in our dataset -- 129,405 (\textasciitilde16\%) documents have affiliation information for at least one author. Therefore, as an approximation, we use information about publications' repositories instead, i.e. we assume authors deposit publications into repositories of institutions they are affiliated with.

There are 728 unique repositories in the dataset, each publication was deposited into 1.16 repositories on average and the largest number of repositories per publication is 31. On the other hand, there are on average 1,286 publications per repository, while 315 repositories contain less than 100 publications and 255 less than 50. Appendix \ref{appendix:map}, Table \ref{tab:largest_repos} presents the ten largest repositories.

\subsection{Country distribution}
\label{sec:data:country_distrib}

\begin{figure}
  \includegraphics[width=0.9\linewidth]{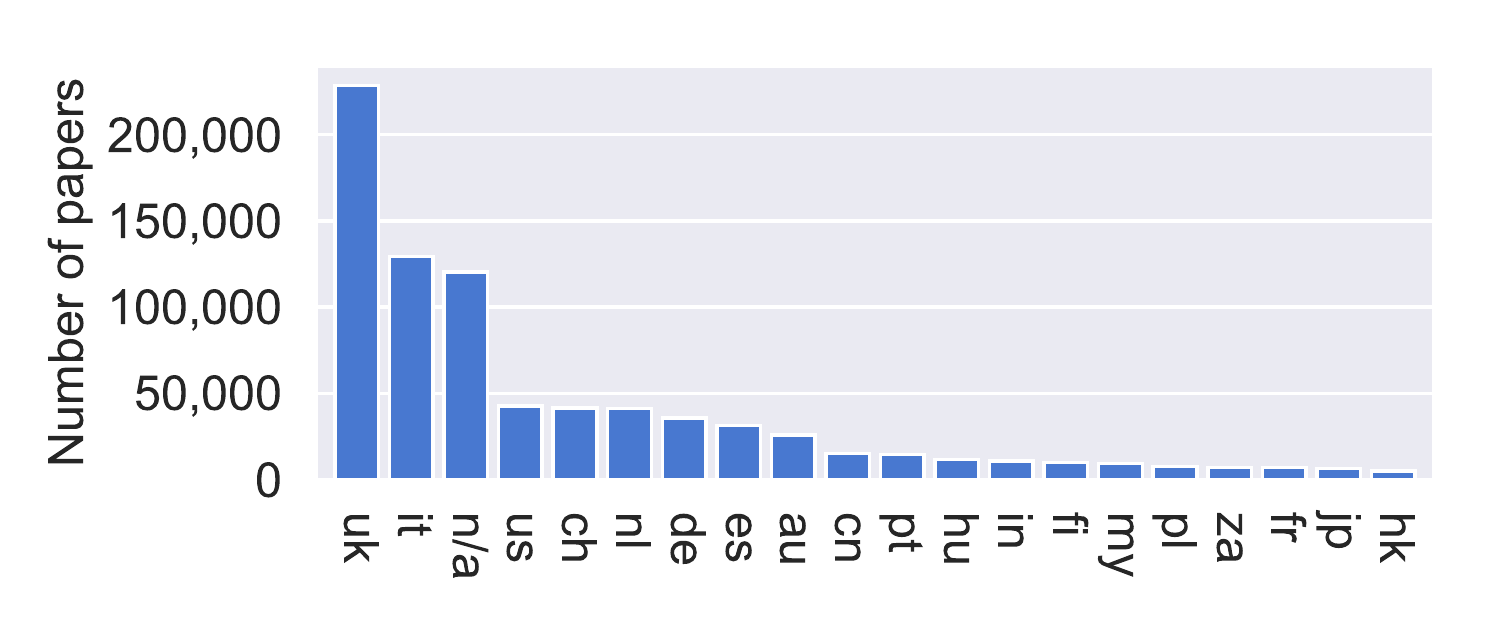}
  \caption{Country distribution of publications in our dataset. The column labelled ``n/a'' represents publications deposited in repositories without a country code (e.g. ArXiv).}
  \label{fig:publication_country}
\end{figure}

To assign publications to countries we use information about repository locations. Figure \ref{fig:publication_country} shows the distribution of publications per country for top 20 countries. Publications affiliated with multiple countries are represented as a full publication for each country (instead of counting only the relevant fraction of the publication).

There are several possible reasons why a large number of publications in our dataset are from the UK. Firstly, the UK had a leading role in the adoption and implementation of repositories comparing to other countries. 
Furthermore, depositing into a repository is now a requirement included in the REF 2021 OA Policy.

\subsection{Date of publication}

\begin{figure}
  \includegraphics[width=0.9\linewidth]{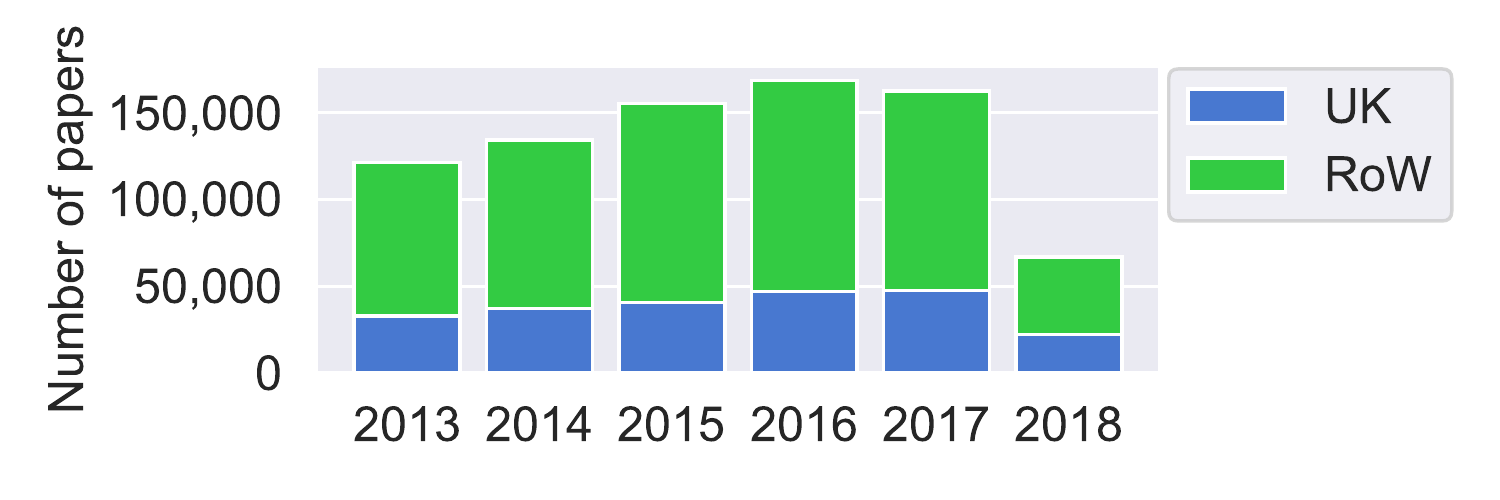}
  \caption{Age of publications in our dataset. Publications with at least one author affiliated with a UK institution are shown in blue, while publications without a UK-based author (labelled ``rest of the world'' -- RoW -- in the figure) are shown in green.}
  \label{fig:publication_year}
\end{figure}

In all experiments we use the date of publication we obtained from Crossref instead of using the date of publication from CORE, as Crossref metadata typically contains more detailed information (e.g. year, month, and day vs. just year). Figure \ref{fig:publication_year} shows the age of publications in our dataset.

As part of our study we are interested in analysing deposit time lag in the UK with regard to the UK OA policy. To understand how many publications in our dataset are from the UK, we distinguish them in the figure by colour -- blue colour represents UK publications, while green colour represent all other publications. 

The drop in publication count in 2018 is due to us not having data for the complete year (we collected data from Crossref in May 2018). The drop in 2017 is likely caused by late deposits -- it is possible that some publications from 2017 had not been deposited yet due to looser policy requirements, authors forgetting to deposit, publisher embargoes, etc.

\subsection{Subject distribution}
\label{sec:data:subject_distrib}

\begin{figure}
  \includegraphics[width=\linewidth]{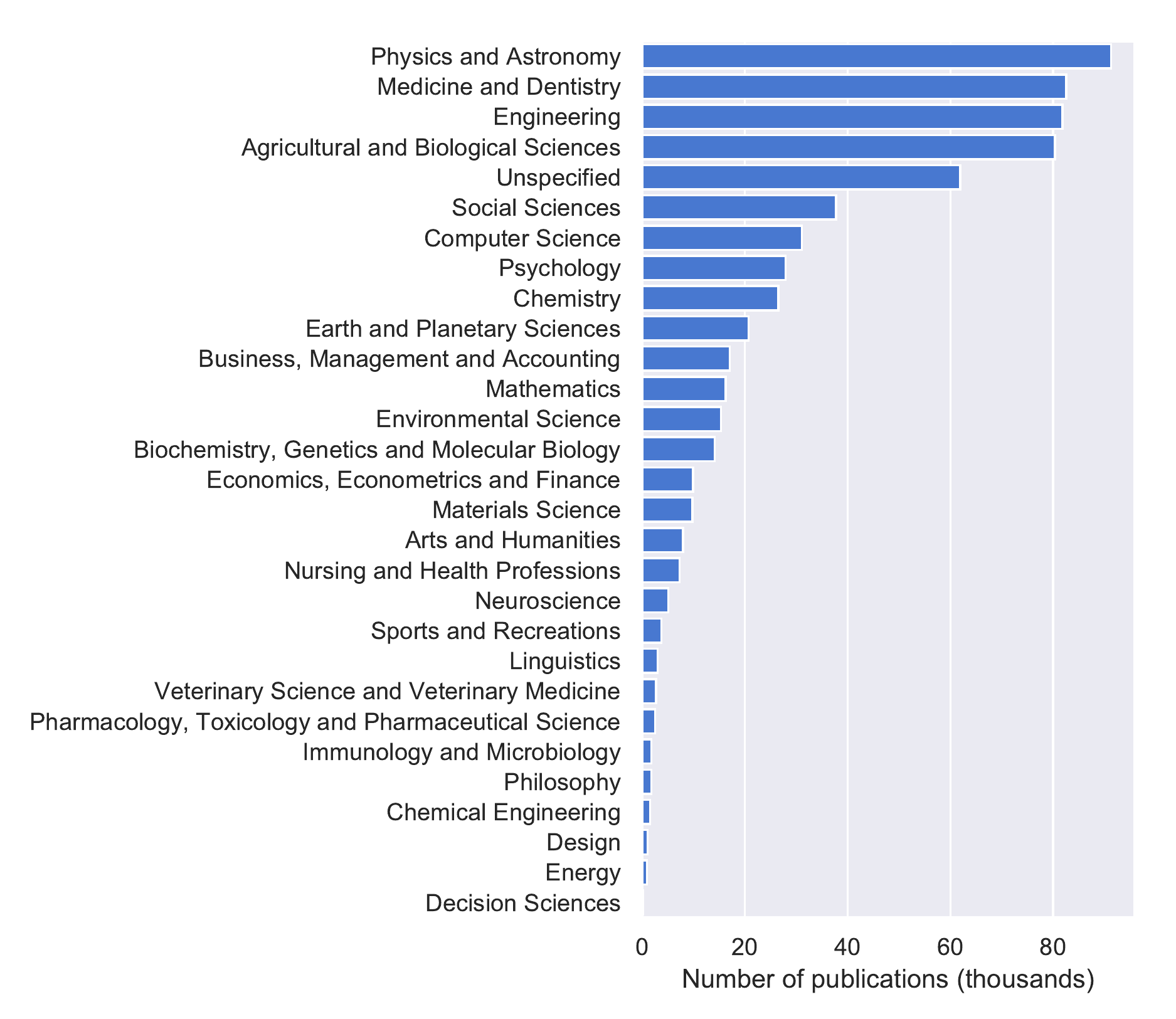}
  \caption{Subject distribution of publications in our dataset.}
  \label{fig:subjects}
\end{figure}

Figure \ref{fig:subjects} shows subject distribution of publications in our dataset. For publications with multiple subjects we only counted the relevant proportion towards each subject. For example, a publication assigned to two subjects is counted as 0.5 towards each subject.

The subjects were obtained from Mendeley in the following way. We used Crossref DOIs to query the Mendeley API\footnote{\url{https://dev.mendeley.com/}} to obtain the metadata Mendeley stores for each article. This metadata contains information about how many readers from each of Mendeley's 28 subjects saved each article in their Mendeley library. Each article was then tagged with the subject in which it accumulated the most readers -- e.g. if an article was read by 20 people in ``Medicine and Dentistry'' and by 5 people in ``Immunology'', we would tag the article with the subject ``Medicine and Dentistry''. In case multiple subjects had the same number of readers the article was tagged with all of those subjects. According to \cite{haunschild2016normalization}, reader counts in Mendeley tend to be skewed towards certain disciplines. The obtained subjects are therefore only an approximation.


We were able to obtain Mendeley metadata for 664,277 publications (\textasciitilde82\%). There are 19 readers per publication on average. Using our subject tagging method described above, 86,731 documents were tagged with multiple subjects (\textasciitilde11\%). Out of those, 65,419 were tagged with two subjects (75\%) and 15,390 with three subjects (18\%), while the rest (5,922, or 7\%) was tagged with between four and ten subjects. While these numbers are lower than existing estimates of the proportion of interdisciplinary research \cite{van2015interdisciplinary}, this could be due to our tagging method.

Additionally, we manually assigned each of the Mendeley subject categories to one of the four REF 2021 Main Assessment Panels\footnote{\url{https://www.ref.ac.uk/about/uoa/}}. These panels are ``A: Medicine, health and life sciences'', ``B: Physical sciences, engineering and mathematics'', ``C: Social sciences'', and ``D: Arts and humanities''. The mapping between Mendeley subjects and REF 2021 panels we used is shown in Appendix \ref{appendix:map}, Table \ref{tab:map}. Figure \ref{fig:ref_panels} shows a distribution of UK publications in our dataset between the four REF 2021 assessment panels.

\begin{figure}
  \includegraphics[width=0.6\linewidth]{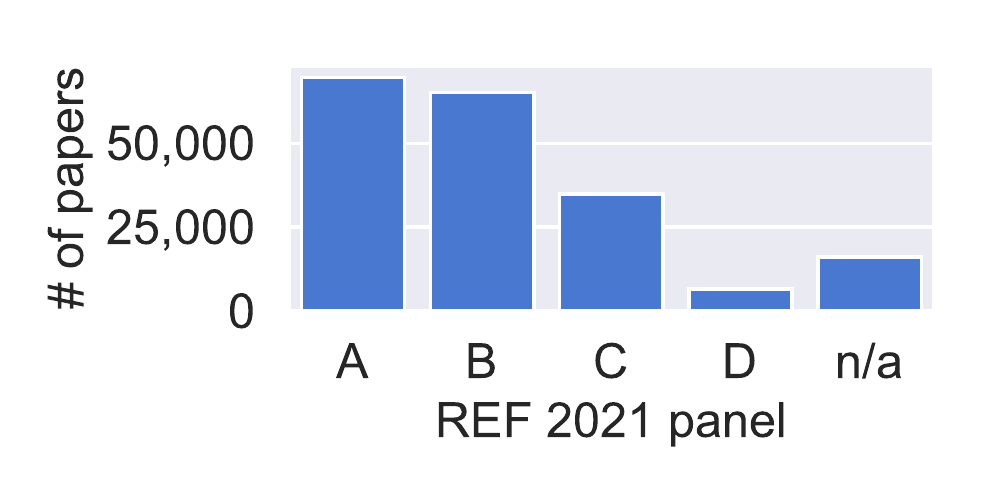}
  \caption{Distribution of UK publications in our dataset into the four main REF 2021 assessment panels.}
  \label{fig:ref_panels}
\end{figure}

\subsection{Crossref acceptance date}
\label{sec:data:acceptance_date}

Crossref metadata contains an \textit{accepted} field which, according to the Crossref API documentation\footnote{\url{https://github.com/Crossref/rest-api-doc/blob/master/api_format.md}}, contains ``date on which a work was accepted, after being submitted, during a submission process''. We have analysed this field for the 800 thousand articles in our dataset. However, we found only 975 articles with the date of acceptance populated. Additionally, for 684 (70\%) this date was the same as the date of publication and for 272 (28\%) the date of acceptance was a later date than the date of publication, showing that the date of acceptance in Crossref is in 99.9\% of cases not available and in 98\% of cases where it is available, it is incorrect. Therefore, we won't use this date in further analysis.

\subsection{ISSN}

As the REF 2021 Open Access Policy applies only to publications with an ISSN, we have included Crossref ISSN numbers in our dataset. We found that 55,014 publications do not have an ISSN number, 12,463 of those are from a UK institution. In our analysis of compliance with the REF 2021 OA Policy have excluded these 12 thousand publications as the policy does not apply to them.

%% file: sections/05_experiments.tex
To calculate deposit time lag for publications in our dataset, we subtracted dates of publication from deposit dates and expressed the difference in days. As a result, negative values mean an article was deposited before being published and positive values mean it was deposited after being published. A histogram of deposit time lag for all publications in our dataset is shown in Appendix \ref{sec:appendix:results}, Figure \ref{fig:dtl_all}.

\subsection{Deposit time lag per country}

\begin{figure}
  \includegraphics[width=0.85\linewidth]{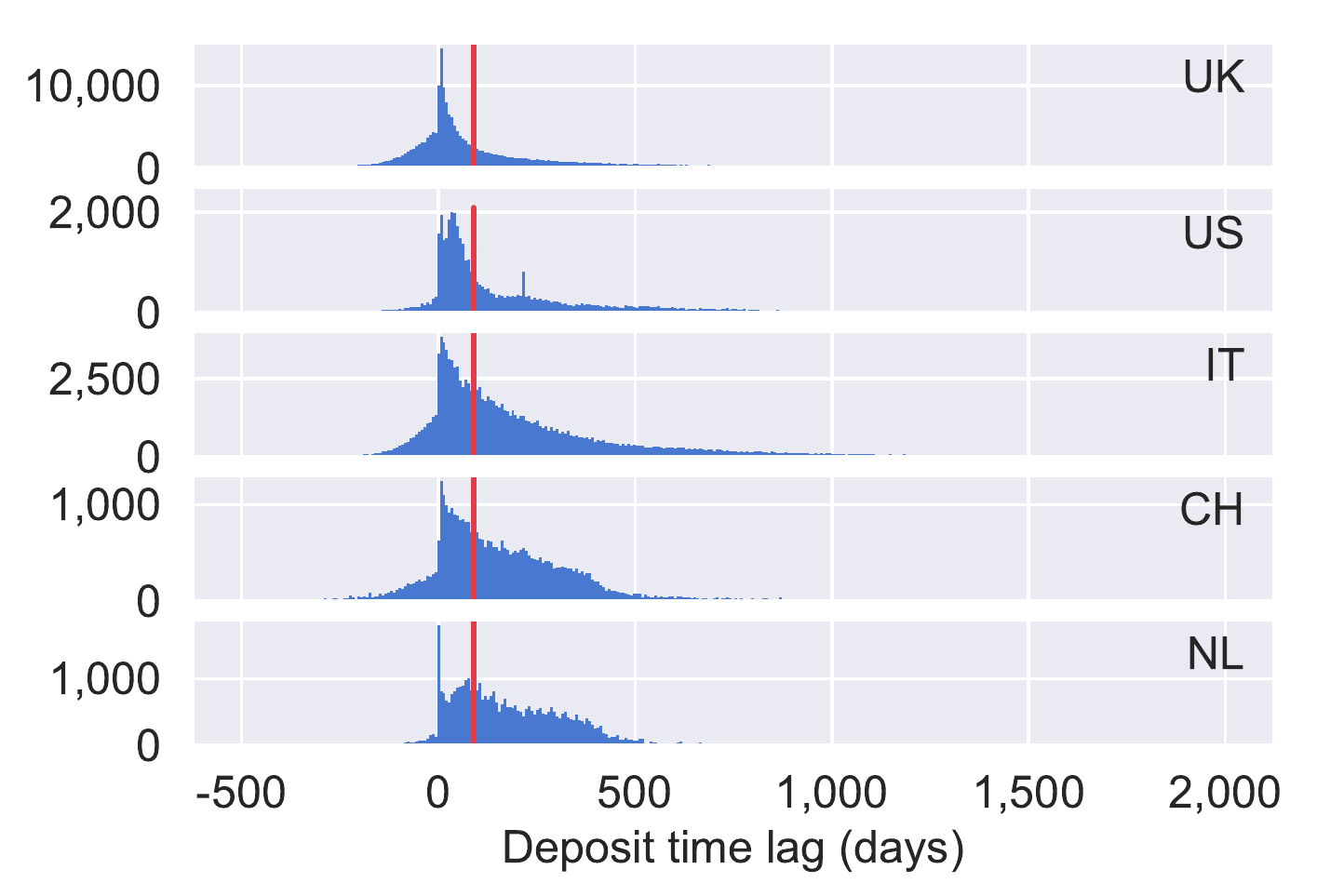}
  \caption{Overall deposit time lag for five countries with the most publications in our dataset. Each bar in the histogram represents one week. The vertical red lines represent 3 months after the date of publication.}
  \label{fig:dtl_country}
\end{figure}

Figure \ref{fig:dtl_country} reveals significant differences in deposit time lag between five countries with the highest number of publications in our dataset. UK publications appear to have the shortest deposit time lag of all five countries, with a large number of articles deposited before or at the time of publication. US publications display a similar pattern, however, deposit time lag in the US peaks a few weeks after publication. On the other hand, Italy, Switzerland, and the Netherlands show a long-tail distribution where deposits peak at the time of publication but decreases slower than in the case of the UK and the US. Furthermore, a large proportion of publications from these countries is deposited with long delays.

Next, we wanted to compare how deposit time lag in these countries has changed over time. One way of doing this is by using all data available to us to calculate average deposit time lag per country and year. This approach has limitations we will illustrate in the following example. Consider deposit dates present in our dataset for articles published in 2013 and in 2017. While articles published in 2013 had just over six years during which they could have been deposited in a repository (our dataset goes until early 2019), publications from 2017 had, in contrast, much shorter time to appear in a repository. It is possible some publications from both years have not been deposited yet, but this is more likely for publications from 2017. This affects yearly deposit time lag in a way which slightly underestimates (decreases) deposit time lag for all publication years, but especially for \textbf{newer} publications.

Another option is to use maximum limit on deposit time lag and filter out all publications which were deposited later than within a specified time frame. To give an example, consider limiting deposit time lag to one year. In this case, only publications from 2013 that were deposited within a year of their publication date (but none of the publications deposited later) would be compared to the same set from 2017. 
This affects yearly deposit time lag in a way which slightly underestimates (decreases) deposit time lag for all years, but especially for \textbf{older} publications, due late deposits becoming less common over time.

As we are not aware of a better way to compare deposit time lag across years that would alleviate the limitations of both of the above mentioned approaches at the same time, we use both approaches in conjunction.

\begin{figure}
  \includegraphics[width=0.8\linewidth]{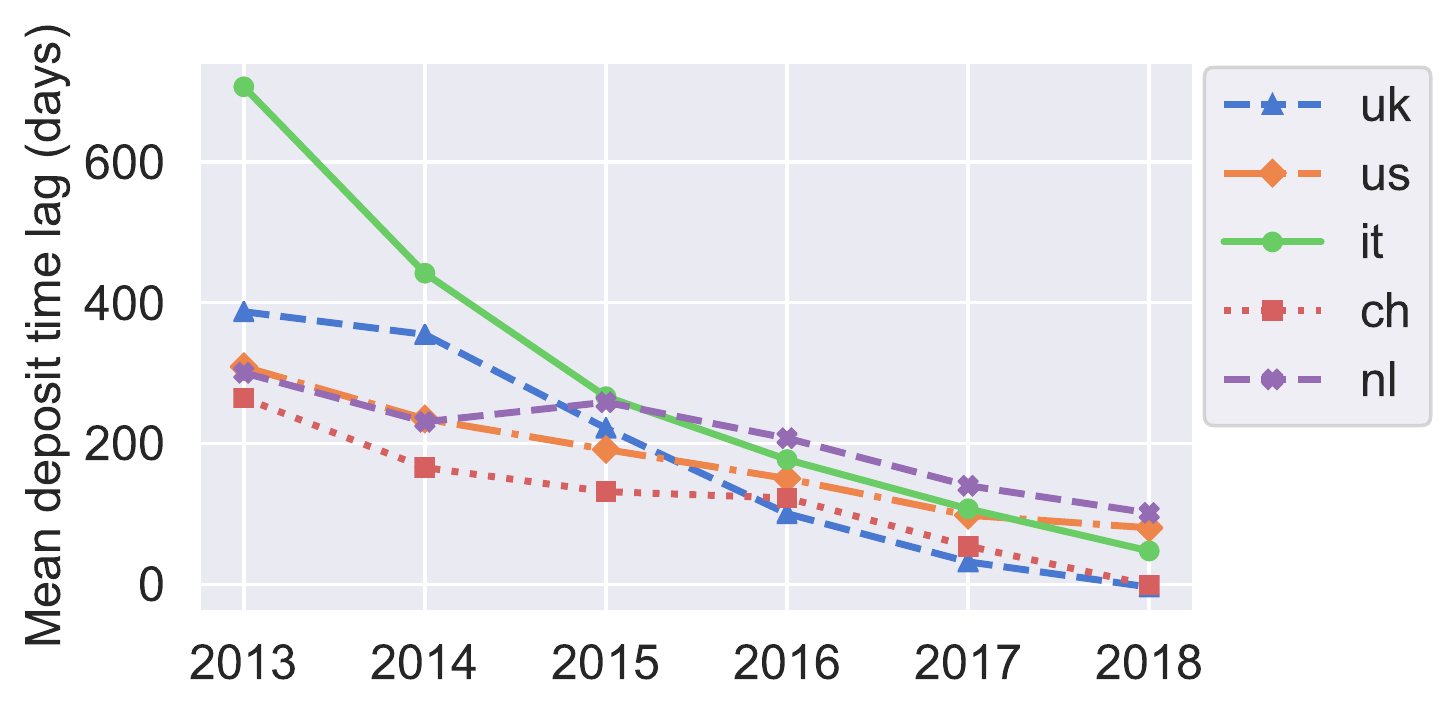}
  \caption{Average deposit time lag per year for five countries with the most publications in our dataset. Figure was created using all available data.}
  \label{fig:dtl_agg}
\end{figure}

\begin{figure}
  \includegraphics[width=0.8\linewidth]{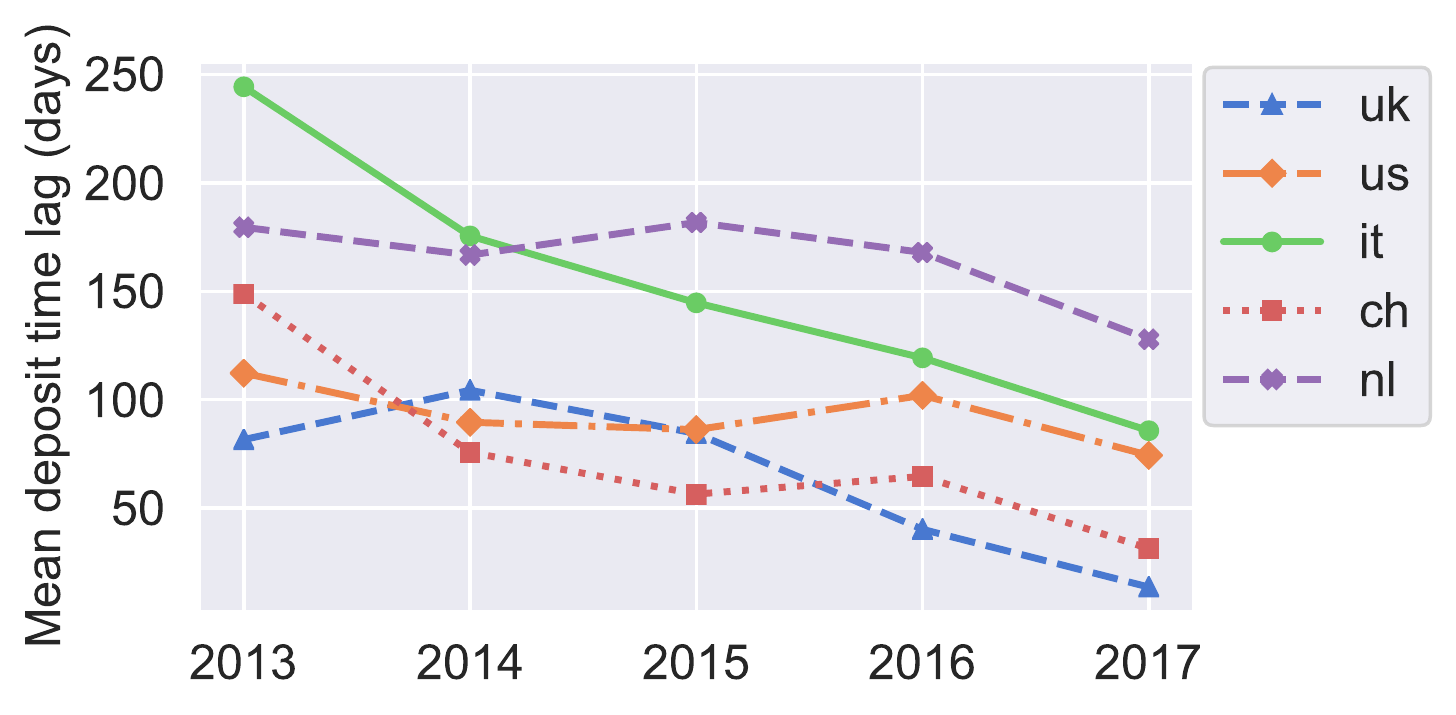}
  \caption{Average deposit time lag per year for five countries with the most publications in our dataset. Figure was created by filtering out all publications which were deposited after a year of being published.}
  \label{fig:dtl_agg_limit}
\end{figure}

Figures \ref{fig:dtl_agg} and \ref{fig:dtl_agg_limit} show average deposit time lag per year and country. In the case of Figure \ref{fig:dtl_agg}, the deposit time lag was calculated using all available data, while in the case of Figure \ref{fig:dtl_agg_limit} it was calculated using one year maximum deposit time lag limit. In the case of Figure \ref{fig:dtl_agg_limit}, year 2018 was excluded as we do not have a complete year of data for it. An additional figure created by applying a maximum deposit time lag limit of two years is shown in Appendix \ref{sec:appendix:results}, Figure~\ref{fig:dtl_agg_limit_2y}.

The figures reveal several interesting trends. Since 2016, the deposit time lag of UK publications is the lowest of all five countries and is negative in Figure \ref{fig:dtl_agg} in 2018 (-3.69 days). In fact, this has not always been the case and, when considering all data including late deposits (Figure \ref{fig:dtl_agg}), the UK was fourth of the selected five countries in 2013 and 2014. Interestingly, this change in average deposit time lag in the UK coincides with the introduction of the REF 2021 OA Policy in 2014. When considering only publications deposited within a year (Figure \ref{fig:dtl_agg_limit}), the UK started as the first of the selected five countries, however, its average deposit time lag had increased in 2014. A possible explanation is the introduction of the REF 2021 OA policy, where researchers started shifting their deposit habits to comply with the policy and as a result deposit more often, but it took time for this shift to become a common practice.


There has been a decreasing trend in deposit time lag for all countries, particularly since 2016. Italy has seen the largest decrease in average deposit time lag from 706 days in 2013 to 48 days in 2018 in the case of Figure \ref{fig:dtl_agg}, and from 244 in 2013 to 86 in 2017 in the case of Figure \ref{fig:dtl_agg_limit}. In 2013, the Italian government passed legislation requiring all research in which at least 50\% of funding was public funding to be made OA \cite{oa_italy}. While we are not aware of any specific deposit time frames associated with this requirement, it is possible it affected deposit practice.

\begin{figure}
  \includegraphics[width=\linewidth]{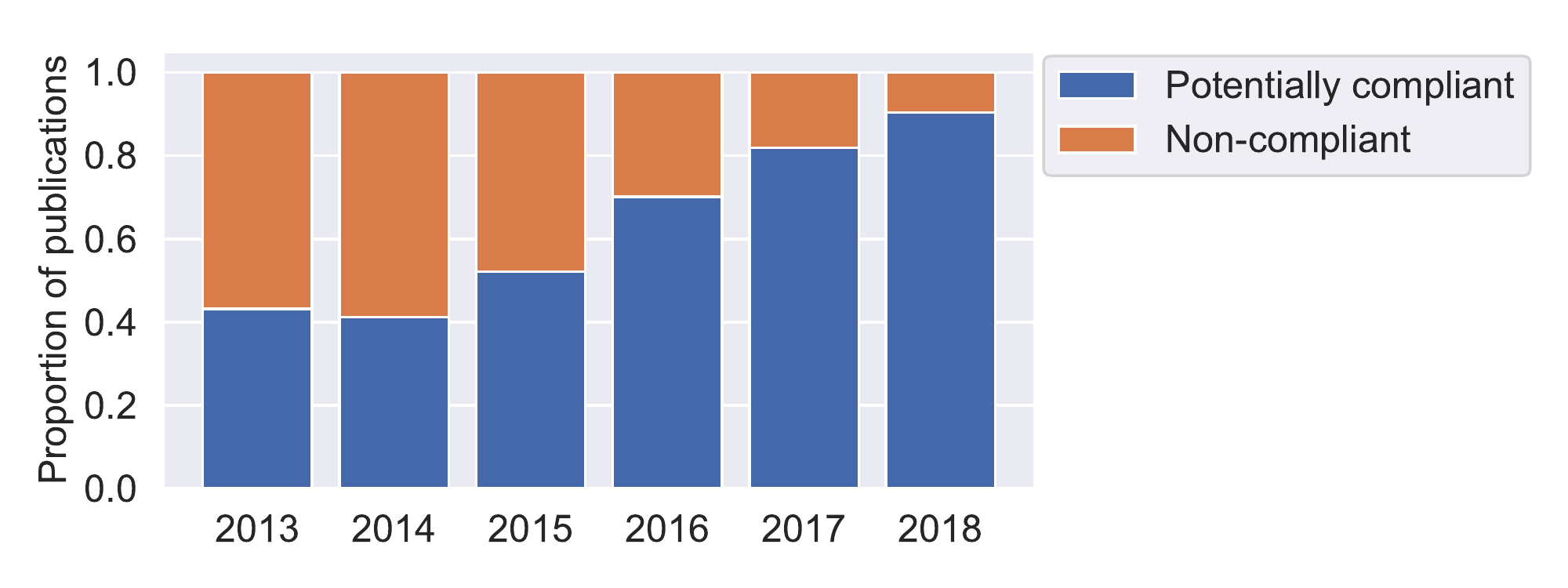}
  \caption{Proportion of non-compliant and potentially compliant UK publications per year.}
  \label{fig:compliance_uk}
\end{figure}

Finally, we analyse deposit time lag with respect to the UK REF 2021 Open Access Policy. To do this, we assign each UK publication to one of the two compliance categories described in Section \ref{sec:methodology:compliance_categories}: ``definitely non-compliant'' -- publications with deposit time lag of more than 90 days, and ``likely compliant'' -- publications with deposit time lag with 90 days or less. The proportion of publications belonging to each category per year is shown in Figure \ref{fig:compliance_uk}.

The figure shows that prior to the REF 2021 OA Policy taking effect in 2016, more than 50\% of publications each year were deposited later than three months after the date of publication. However, the situation has changed after the policy took effect in April 2016. In 2017, 80\% of papers were made available in an OA repository within three months of the date of publication, or even earlier. While we do not yet have complete data for 2018 (our sample contains data until May 2018), we can observe that compliance is still 
increasing.

\subsection{Deposit time lag per repository}
\label{sec:dtl:repository}

\begin{figure}
  \includegraphics[width=0.9\linewidth]{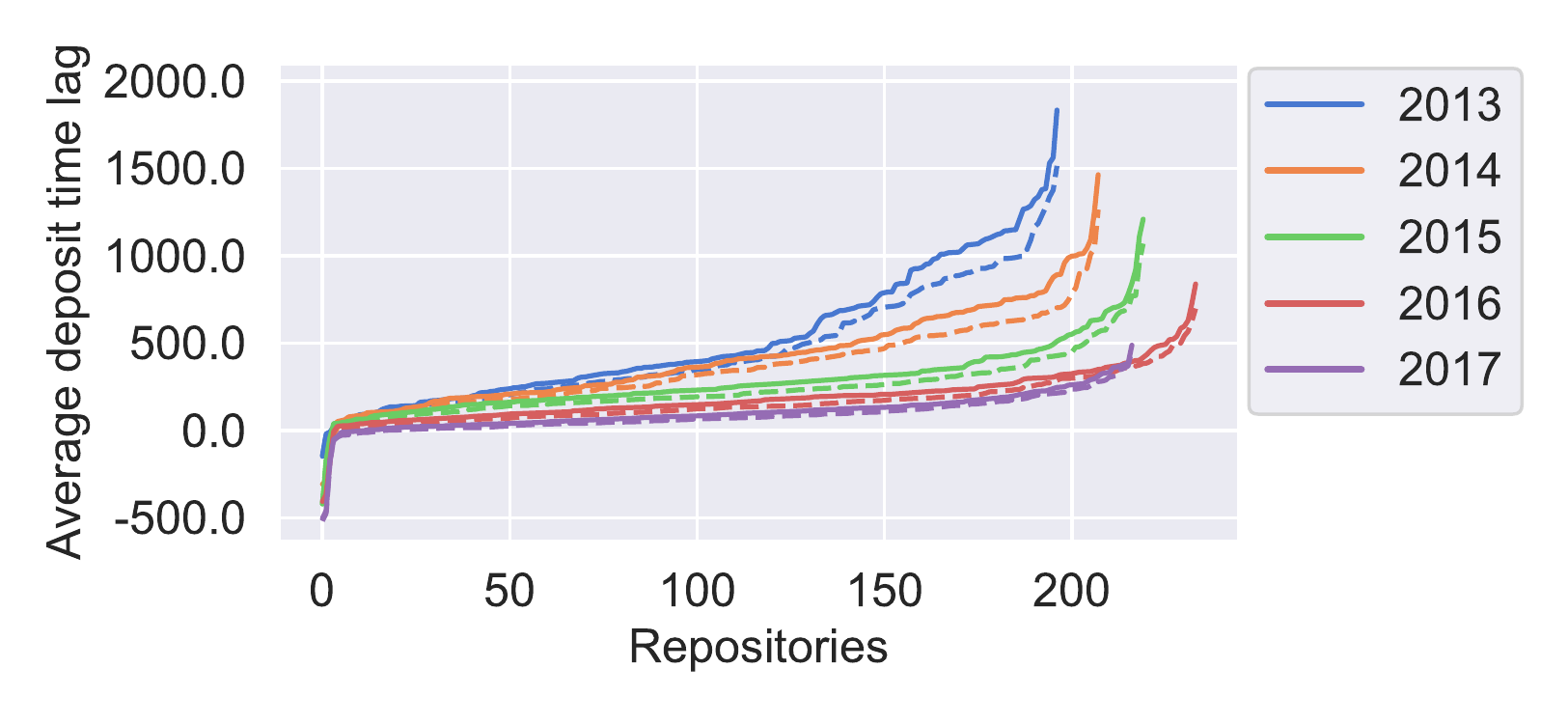}
  \caption{Deposit time lag per repository and year. The full lines show ``single repository deposit time lag'' and the dashed lines show ``any repository deposit time lag''.}
  \label{fig:repo_dtl}
\end{figure}

Our next question is whether there is a difference between deposit time lag of different repositories and how this has changed over time. Figure \ref{fig:repo_dtl} shows deposit time lag per year for all repositories with more than 100 publications in a given year. To produce this figure, we have calculated the following two statistics for each repository:

\begin{enumerate}
    \item \textbf{Single repository deposit time lag.} Deposit time lag with respect to the publications' deposit date in a \textbf{given} repository. In this case, we do not take into account that a publication may have been deposited into multiple repositories. For example, if a publication was deposited into the University of Cambridge repository, we only consider the date of deposit into this repository.
    \item \textbf{Any repository deposit time lag.} Deposit time lag calculated with respect to the publications' deposit date in \textbf{any} repository. For example, if a publication was deposited into the University of Cambridge repository as well as elsewhere, we simply use the first of the two dates to calculate deposit time lag.
\end{enumerate}

To produce the full lines in Figure \ref{fig:repo_dtl}, we have sorted the repositories according to their ``single repository deposit time lag'' values from the lowest to the highest. The dashed lines were produced the same way, but using the ``any repository deposit time lag'' values. The figure reveals significant differences between repositories, which have reduced over time, but remain high. For 2013 publications, the difference between the repository with the lowest and the highest ``single repository deposit time lag'' was 1,982 days, and the standard deviation across all repositories was 377 days. In 2017, these numbers have dropped to 991 days and a standard deviation of 108. The figure also reveals that by aggregating data from all repositories, the deposit time lag can be lowered.

\begin{figure}
  \includegraphics[width=0.8\linewidth]{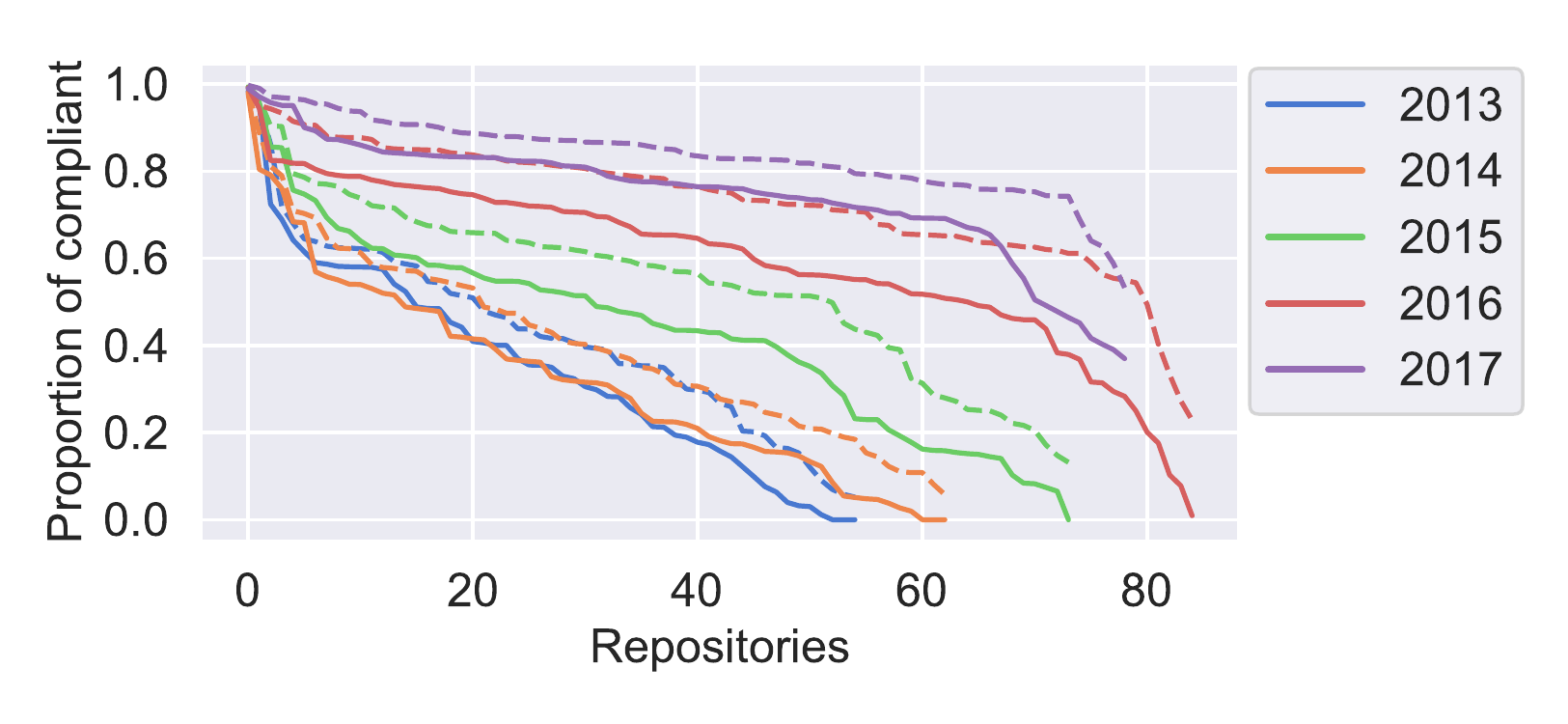}
  \caption{Proportion of likely compliant publications per repository and year. The full and dashed lines show ``\textbf{single}'' and ``\textbf{any}'' repository compliance, respectively.}
  \label{fig:repo_compliance}
\end{figure}

We have produced a similar figure for UK repositories showing the proportion of ``likely compliant'' publications per repository. Similarly to Figure \ref{fig:repo_dtl}, Figure \ref{fig:repo_compliance} was produced by calculating two statistics for each repository: \textbf{single repository compliance} (full lines), i.e. proportion of likely compliant publications when considering deposits only in a \textbf{single} repository, and \textbf{any repository compliance} (dashed lines), i.e. proportion of likely compliant publications with respect to their deposit date in \textbf{any} repository. In both cases, the repositories were sorted from the most to the least compliant. It can be seen that repository compliance has increased rapidly from 2014 onward, particularly between 2015 and 2016. As the UK REF 2021 OA Policy was introduced in 2014, it may be one of the reasons for this increase. The figure also shows that aggregating research outputs from multiple repositories may help improve repository compliance.

\subsection{Deposit time lag per subject}

\begin{figure}
  \includegraphics[width=\linewidth]{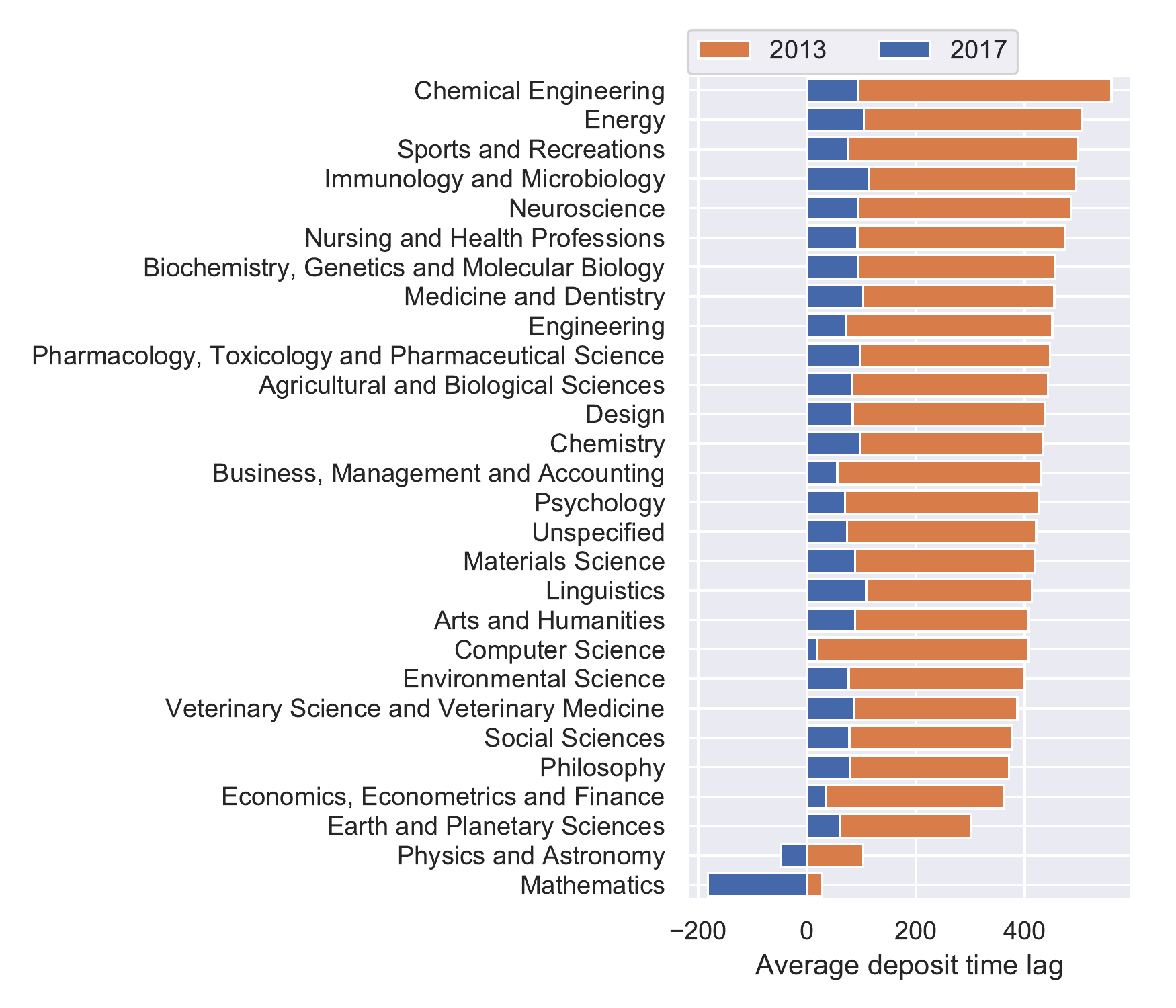}
  \caption{Average deposit time lag per subject in 2013 and 2017. The bars in the figure are not stacked but instead placed on top of each other, i.e. the bars of both years have the same baseline of zero.}
  \label{fig:dtl_subj}
\end{figure}

Finally, we investigated whether there were any differences in deposit time lag between different subjects. Figure \ref{fig:dtl_subj} shows average deposit time lag per subject in 2013 and 2017. To produce this figure we have removed a single subject (Decision Sciences) with less than 100 publications in one year. The figure shows that while there were significant differences between subjects in 2013, these were largely diminished by 2017. The figure also reveals smaller differences between subjects than the differences observed between repositories shown in Figure \ref{fig:repo_dtl}. In 2013, the difference between the highest and the lowest average deposit time lag per subject was 532 days and standard deviation across all subjects was 107 days. In 2017 the range was 295 and standard deviation was 57 days.


On the other hand as we have shown in Section \ref{sec:dtl:repository}, range and standard deviation across all repositories were 1,982 and 377 in 2013, and 991 and 108 in 2017. If we consider only publications from a single subject, the differences between repositories remain high. For example, using only publications from ``Physics and Astronomy'' (our largest subject), range and standard deviation were 1,787 and 370 in 2013, and 940 and 174 in 2017. The situation is similar for other subjects. This suggests institutional policies, particularly when harmonised with funder policies, may be stronger drivers of OA than disciplinary culture.

\begin{figure}
  \includegraphics[width=0.75\linewidth]{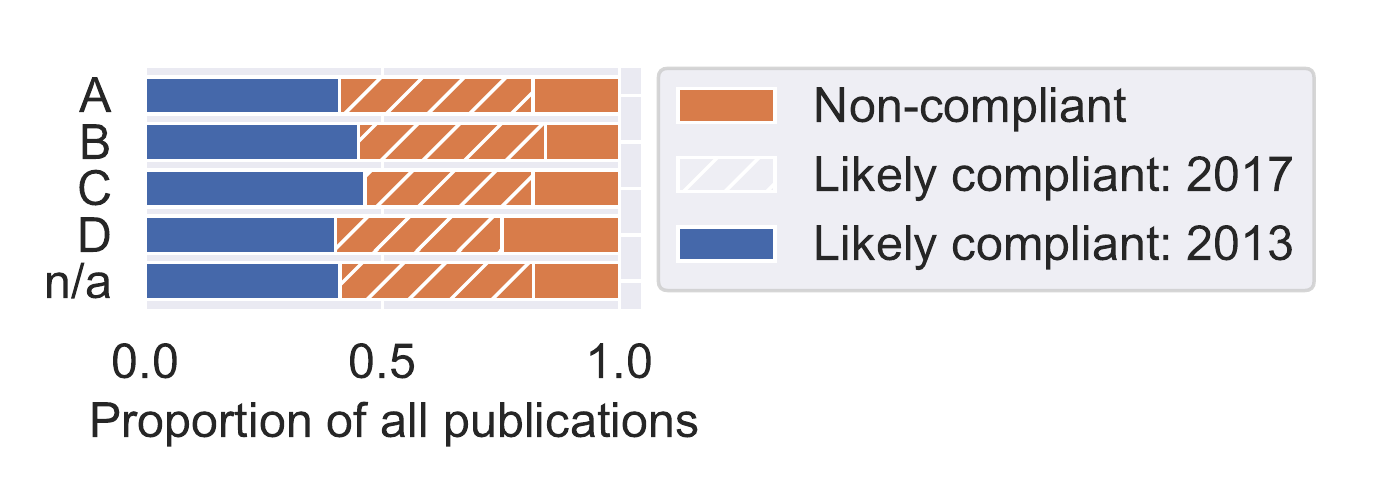}
  \caption{Proportion of likely compliant and non-compliant publications per each of the main REF 2021 assessment panels.}
  \label{fig:compliance_panel}
\end{figure}

Finally, Figure \ref{fig:compliance_panel} shows the proportion of likely-compliant and non-compliant publications across the four main REF 2021 assessment panels (Section \ref{sec:data:subject_distrib}) in 2013 and in 2017. The figure shows there has been significant increase in compliance over the five year period, which has been similar across all four panels.

%% file: sections/06_discussion.tex
Our findings indicate that deposit time lag has been decreasing globally. 
However, we have observed major differences in deposit time lag across institutions and significant differences between subjects. Furthermore, we have shown that the deposit time lag has been shortening over the last 5 years both globally and in the UK. Our results suggest that the REF 2021 OA Policy likely helped to reduce deposit time lag. 
The results outlined in this paper present a preliminary study of deposit time lag and compliance with existing OA policies. There are many areas where this study could be enhanced and broadened. 

The matching of articles between Crossref and CORE was done by means of the articles' metadata (titles, years of publication, and first author names). This is a strict approach that may result in lower recall due to minor differences in metadata, such as listing authors in incorrect order, typos, differences in punctuation, etc. While our present study has been precision oriented, i.e. our aim was to produce as clean data as possible, in the future, we would like to improve our recall. This would also allow us to study deposit rates, i.e. the proportion of articles that get deposited into OA repositories compared to articles that do not, in addition to deposit time lag. Improving our recall could be done in a number of ways. For example, in addition to the metadata we already use for the matching, we could utilise all other metadata available to us, such as abstracts, and employ looser matching techniques such as those used in article deduplication \cite{jiang_2014_rule}.


For this initial study we make the assumption that if a metadata record is in the repository, the full text is also deposited. This is because validating if the full text is deposited is a complicated process which is outside of the scope of this work. The OAI-PMH protocol does not guarantee a link to the publication full text will be in the metadata even if the full text was deposited into the repository. To check if an article full text was deposited, we would have to crawl all links provided in the OAI-PMH metadata and correctly match the identified documents to the publication metadata. Therefore, as our present study focuses on deposit time lag rather than presence of the full text, we decided not to perform this check. 

As our analysis relies on deposit dates, publications that have never been deposited into a repository are not included in our study. Consequently, this means that the proportion of publications that are potentially compliant with the REF 2021 OA Policy are compared against non-compliant but deposited publications, rather than all publications. To quantify missing deposits, we would have to be able to correctly match all CORE publications to their Crossref metadata. This is out of the scope of our study, as the focus of our study is on deposit time lag rather than the analysis of the proportion of missing deposits. However, to allow for as many publications to be included in our study we have collected deposit dates almost a year (in March 2019) after collecting publication metadata (May 2018).

%% file: sections/07_conclusion.tex
The aim of this study was to investigate how much time does it take for authors to deposit their articles in OA repositories in relation to when these articles get published. Furthermore, our goal was to investigate if OA policies might have reduced this time, and if compliance with such policies can be effectively tracked. We collected dates of publication and deposit dates for 800 thousand articles published around the world between 2013 and 2018, and compared the difference between these dates across time, country, subject, and repository.

We have shown that the time between publication and deposit has decreased significantly over the 2013-2017 period globally, by 472 days per country on average across all countries in our dataset. We have also shown that after the introduction of the UK REF 2021 OA Policy, this decrease in the UK has accelerated, and in 2018 the mean difference between publication and deposit dates has become negative (-3.69 days), meaning that, as of early 2018, on average, UK publications potentially become OA immediately or even slightly before publication. The key message of our paper is that this observation supports the argument for the inclusion of a strictly time-limited deposit requirement in OA policies. Furthermore, our work demonstrates that countries which now have a time frame on deposits included in their OA policies can develop reliable tracking mechanisms for monitoring the effects of such policies.

Based on the presented methodology, we have developed a tool for tracking the time lag between article publication and deposit which relies on data from thousands of repositories. We hope the tool will be useful to authors, funders and institutions who intend to improve the accessibility of research and improve compliance with existing OA policies. To support further studies on the deposit of research outputs in OA repositories, we release our dataset of 800 thousand publications and the source codes of our analysis\footnote{\url{http://github.com/oacore/jcdl_2019}}.

%% file: sections/99_appendix.tex
\section{Data preparation and statistics}
\label{appendix:map}

\bgroup
\begin{table}[h!]
\caption{The ten largest repositories in our dataset.}
\label{tab:largest_repos}
\begin{tabular}{|p{\linewidth-2.7cm}|p{2cm}|}
\hline 
Name & Publications \\ 
\hline
\hline
ArXiv e-Print Archive & 97,594 \\ \hline
White Rose Research Online & 24,019 \\ \hline
ZORA & 20,617 \\ \hline
Utrecht University Repository & 20,304 \\ \hline
Enlighten & 19,267 \\ \hline
Radboud Repository & 17,837 \\ \hline
ZENODO & 17,100 \\ \hline
Universit\`{a} di Roma La Sapienza Repository & 14,795 \\ \hline
Online Research @ Cardiff & 14,261 \\ \hline
Universit\`{a} di Padova Repository & 14,077 \\
\hline 
\end{tabular}
\end{table}
\egroup

\bgroup
\begin{table}[h!]
\caption{Mapping of Mendeley subjects to REF 2021 Main Panels.}
\label{tab:map}
\begin{tabular}{|p{\linewidth - 2.7cm}|p{2cm}|}
\hline 
Mendeley subject & REF Main Panel \\ 
\hline
\hline
Agricultural and Biological Sciences & A \\ \hline
Arts and Humanities & D \\ \hline
Biochemistry, Genetics and Molecular Biology & A \\ \hline
Business, Management and Accounting & C \\ \hline
Chemical Engineering & B \\ \hline
Chemistry & B \\ \hline
Computer Science & B \\ \hline
Decision Sciences & C \\ \hline
Design & D \\ \hline
Earth and Planetary Sciences & B \\ \hline
Economics, Econometrics and Finance & C \\ \hline
Energy & B \\ \hline
Engineering & B \\ \hline
Environmental Science & B \\ \hline
Immunology and Microbiology & A \\ \hline
Linguistics & D \\ \hline
Materials Science & B \\ \hline
Mathematics & B \\ \hline
Medicine and Dentistry & A \\ \hline
Neuroscience & A \\ \hline
Nursing and Health Professions & A \\ \hline
Pharmacology, Toxicology and Pharmaceutical Science & A \\ \hline
Philosophy & D \\ \hline
Physics and Astronomy & B \\ \hline
Psychology & A \\ \hline
Social Sciences & C \\ \hline
Sports and Recreations & C \\ \hline
Unspecified & n/a \\ \hline
Veterinary Science and Veterinary Medicine & A \\
\hline 
\end{tabular}
\end{table}
\egroup

\section{Analysis results}
\label{sec:appendix:results}

\begin{figure}
 \includegraphics[width=0.85\linewidth]{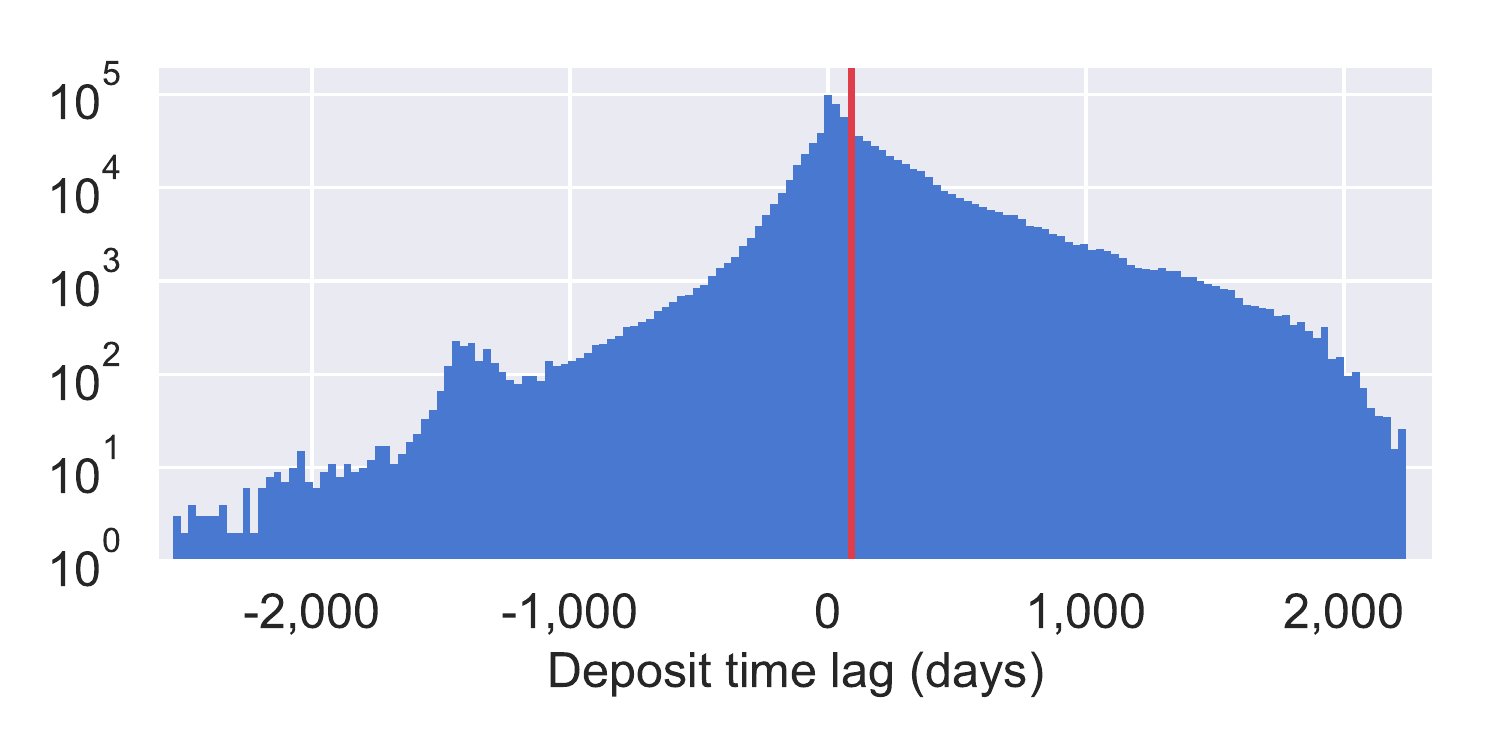}
 \caption{Deposit time lag in days for all publications in our dataset. The histogram was created by aggregating 30 days at a time, i.e. each bar represents one month. The y-axis is logarithmic and the vertical red line represents 3 months after the date of publication.}
 \label{fig:dtl_all}
\end{figure}

Figure \ref{fig:dtl_all} shows a histogram of deposit time lag in days for all publications in our dataset. The vertical red line in the figure represents 3 months after the date of publication which is the cut-off between our ``likely compliant'' and ``definitely non-compliant'' categories (Section \ref{sec:methodology:compliance_categories}), meaning that all publications that fall on the right side of the line would not be compliant with the REF 2021 OA Policy. The maximum deposit time lag in our dataset is 2,241 days. This large time lag is possible, as the earliest date of publication in our dataset is January 1, 2013, while the deposit dates were collected between March 7 and March 18, 2019 -- the difference between January 1, 2013 and March 18, 2019 is 2,268 days. The graph shows a large portion of articles in our dataset was deposited retrospectively many years after publication.

\begin{figure}
  \includegraphics[width=0.8\linewidth]{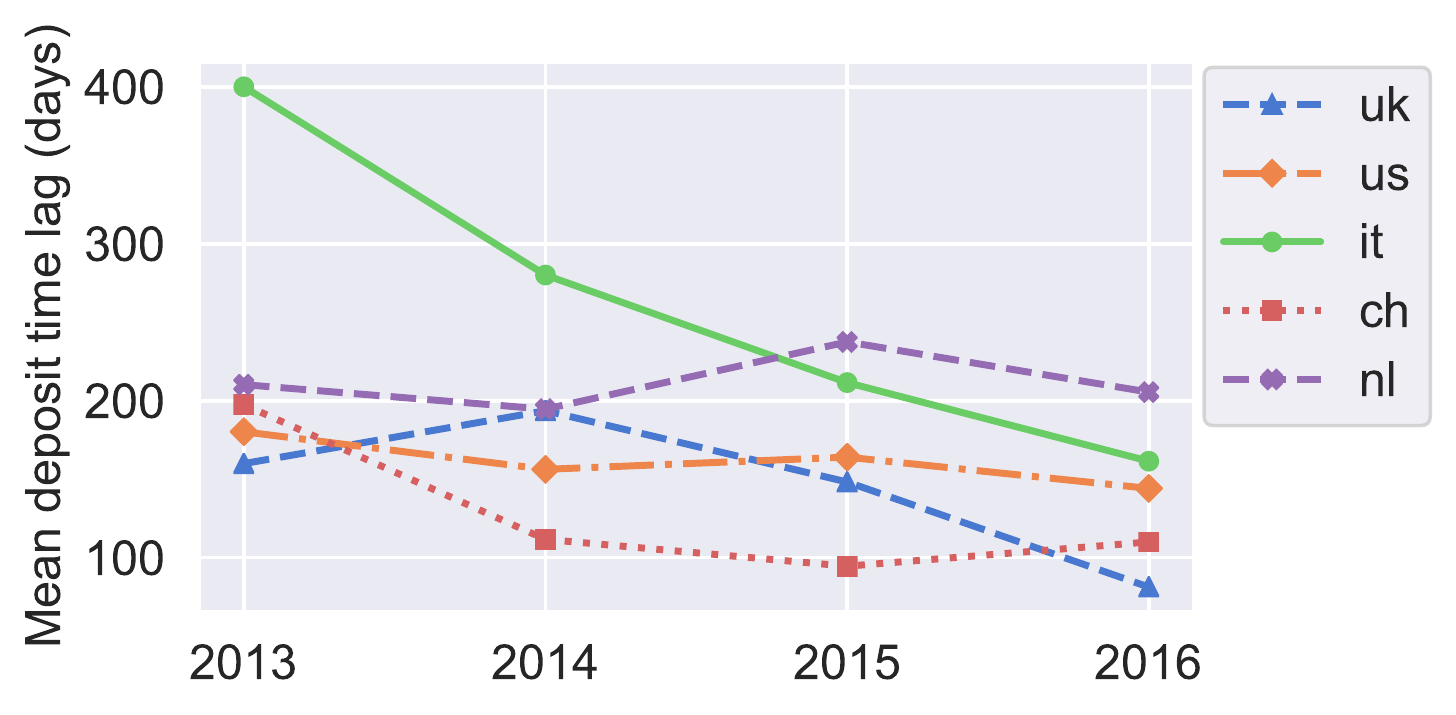}
  \caption{Average deposit time lag per year for five countries with the most publications in our dataset. Figure was created by filtering out all publications which were deposited later than within two years of being published.}
  \label{fig:dtl_agg_limit_2y}
\end{figure}

Figure \ref{fig:dtl_agg_limit_2y} shows average deposit time lag per country and year. To prepare the figure, data was first filtered by removing all publications which were deposited later than two years after being published.

